\UseRawInputEncoding
\documentclass[letterpaper]{article} 
\usepackage{aaai2026}  
\usepackage{times}  
\usepackage{helvet}  
\usepackage{courier}  
\usepackage[hyphens]{url}  
\usepackage{graphicx} 
\usepackage{natbib}  
\usepackage{caption} 
\usepackage{xcolor}
\usepackage{multirow}
\usepackage{subcaption}
\usepackage{pifont}
\usepackage{amsmath}
\usepackage{booktabs}
\usepackage{tcolorbox}
\tcbuselibrary{listings, breakable, skins}

\usepackage{algorithm}
\usepackage{algorithmic}

\usepackage{xcolor}

\usepackage{newfloat}
\usepackage{listings}
\DeclareCaptionStyle{ruled}{labelfont=normalfont,labelsep=colon,strut=off} 
\floatstyle{ruled}
\newfloat{listing}{tb}{lst}{}
\floatname{listing}{Listing}
\nocopyright

\title{How Human Feedback Shapes AI-generated Community Notes}

\author{
    Soham De,
    Isaac Slaughter, 
    Jiawei Guo, 
    Qiao-Yun Cheng, \\
    Jiayuan Yan,
    Sruti Banerjee,
    Martin Saveski
}
\affiliations{
    University of Washington
}

\begin{document}

\maketitle

\begin{abstract}
Community Notes, a bridging-based crowd-sourced fact-checking system, has emerged as a new mechanism for moderating misleading information on social media and has been adopted by major platforms including X, Facebook, Instagram, Threads, and TikTok. Since its introduction, there has been an open question about what role AI could play in scaling and optimizing the system. Recently, X extended its Community Notes system by introducing \textit{Collaborative Notes}: notes initially drafted by an LLM and iteratively refined based on feedback from human contributors. In this work, we systematically analyze the complete corpus of 19,146 collaborative notes and 211,850 instances of human feedback. First, we develop a taxonomy of human suggestions for improving AI-generated note drafts and find that suggestions involving factual corrections and additional context are most likely to be incorporated, while subjective policy judgments rarely are. Second, we examine changes in helpfulness across versions of collaborative notes and find that human feedback leads to more helpful notes, with the greatest impact coming from suggestions that challenge the main claim in the previous draft, particularly when submitted by more active contributors. Finally, we find that although collaborative notes improve through human feedback, they reach helpful status and are shown on the platform at lower rates than human-only or AI-only notes, with limited human participation emerging as a key bottleneck. Nevertheless, rather than serving as a weaker substitute, collaborative notes tend to play a complementary role, predominantly targeting posts that do not attract human-only or AI-only notes. Our analysis provides an initial description of efforts to use AI to improve crowdsourced content moderation in a real-world moderation system and outlines pathways for future improvements to such features.
\end{abstract}

\section{Introduction}
\label{sec:intro}
Community Notes has emerged as the industry-standard mechanism for providing additional context on potentially misleading social media content. In this system, fact-checking is not performed by professional fact-checkers or platforms. Instead, ordinary users of the platform step up to request additional context on potentially misleading posts, propose fact-checking notes and also rate on proposed fact-checking notes, representing a bottom-up approach to content moderation. Originally launched as Birdwatch \cite{wojcik2022birdwatch} on X (then, Twitter) in 2019, the approach has since been adopted in various forms by most major platforms, including YouTube, Facebook, Instagram, Threads and TikTok. Studies investigating Community Notes find that the notes are effective at reducing the engagement with and spread of misinformation when attached to posts~\cite{chuai2026community, slaughter2025community}, but also highlight persistent challenges around coverage~\cite{wack2026laziness} and the speed at which notes are attached to posts~\cite{slaughter2025community}.

Several directions for overcoming these challenges have been proposed in the literature, ranging from algorithmic improvements to the scoring process~\cite{goyal2026quality} to the integration of AI in various parts of the system~\cite{li2025scaling}. These AI-assisted approaches include enhancing existing human-written notes~\cite{de2025supernotes, zhang2025commenotes} as well as developing more autonomous, end-to-end systems for automated note generation~\cite{wu2025beyond, li2026ai}. To retain the system’s credibility and trust, any attempts to scale Community Notes with AI need to be designed in ways that preserve human judgment in determining what the community ultimately finds helpful.

Since the inception of Community Notes, X has made several attempts to incorporate AI into the note-writing workflow. In June 2025, the platform released an API allowing bots to author draft notes on posts flagged as potentially misleading by users. This allows Community Notes contributors to design their own AI-powered fact-checking bots that can automatically generate note drafts, which are then rated by humans before being shown publicly. More recently, in February 2026, X introduced \textit{Collaborative Notes}, an extension of the system in which an LLM (Grok) automatically drafts a community note on flagged content, with the draft iteratively refined through helpfulness ratings and free-text suggestions submitted by Community Notes contributors.

Collaborative human-AI note-writing offers several advantages over human-only note-writing. LLM assistance may produce more thoroughly researched drafts than individual contributors could create on their own, while also incorporating human-provided context that may be critical in fast-evolving situations. It also simplifies the task from writing a note from scratch to suggesting edits, which may lower the barrier to participation and broaden engagement across the contributor community. At the same time, the collaborative process raises new questions: (i) whether the back-and-forth between human suggestions and LLM revisions actually improves note quality, (ii) whether it does so effectively, and (iii) whether the resulting notes are ultimately more helpful than those written through existing workflows.

In this work, we present a systematic analysis of the human-AI collaborative notes on X, aiming to characterize the current state of the system and inform future design decisions. We center our analysis on the following research questions:

\begin{description}
    \item[\textbf{RQ1}] \textbf{What feedback do humans provide on AI-generated notes?} \newline
    Specifically, who are the raters that engage in this system and what kinds of suggestions do they provide? What kinds of suggestions are more likely to be adopted by the LLM as it revises new versions?
    \item[\textbf{RQ2}] \textbf{How do collaborative notes evolve over time?} \newline
    Specifically, do collaborative notes become more or less helpful as they are rewritten based on the human feedback? What drives improvement between versions of collaborative notes?
    \item[\textbf{RQ3}] \textbf{Are collaborative notes effective?} \newline
    Specifically, how do collaborative notes perform compared to non-collaborative human- or AI-written notes on the platform? What are potential explanations for their relative performance?
\end{description}

\noindent
To answer these questions, we collected the complete corpus of all collaborative notes on X up to May 5, 2026 resulting in $19,146$ note versions on $10,600$ unique posts, and  $211,850$ ratings and suggestions, over a span of 3 months. We performed a mixed-methods analysis, using qualitative methods to develop a taxonomy to characterize the suggestions provided by humans and quantitative methods to describe the evolution of collaborative notes over time, and how they compare against non-collaborative notes. In doing so, we make the following contributions:

\begin{enumerate}
     \item A qualitative analysis resulting in a \textbf{taxonomy of the kinds of suggestions} provided by human raters on AI-drafted notes, and the prevalence of each type, including which are most likely to be incorporated into the next version (Section \ref{sec:results_human_input}). We find that participation is dominated by experienced contributors, and that suggestions grounded in factual corrections, added context, or source critique are the most likely to be incorporated, while subjective policy judgments are rarely~incorporated.
    \item An analysis characterizing \textbf{how a collaborative note improves over time}, and how that depends on the overall chain length, kinds of suggestions provided, and the experience and political lean of the raters providing them (Section \ref{sec:results_note_evolve}). We find that while notes generally improve with sustained revision, improvement is rarely monotonic, and that longer revision chains are associated with significantly greater quality gains.
    \item A comparative analysis examining the \textbf{effectiveness of collaborative notes against non-collaborative human- and AI-written notes}, and highlighting potential explanations and opportunities for improvement (Section \ref{sec:results_compare}). We find that collaborative notes currently underperform both baselines on helpfulness and take longer to reach rater consensus, with a lack of ratings emerging as a key structural bottleneck and point to concrete design interventions to address it.
\end{enumerate}

\section{Related Works}
\label{sec:literature_review}

Our research questions and analyses draw from a rich body of existing research on crowd-sourced fact-checking, including how human-AI collaboration can aid such processes. Below, we highlight work most closely related to this study.

\subsection{Crowd-sourced fact-checking on social media}

Crowd-sourced fact-checking emerged as a promising approach to curbing misinformation on social media \cite{slaughter2025community}. The underlying intuition---that aggregated lay judgments can effectively surface low-quality information---was explored in early experiments \cite{florin2010crowdsourced} and later validated at scale, showing that crowd-sourced assessments of news source quality can closely match expert judgments \cite{pennycook2019fighting, allen2021scaling, resnick2023searching}. Community Notes---Twitter/X's crowd-sourced annotation system---has emerged as one of the most studied implementations of this paradigm \cite{prollochs2022community}, with subsequent work examining how notes diffuse through the network \cite{drolsbach2023diffusion}, how the crowd selects fact-checking targets \cite{pilarski2024community}, how notes increase user trust \cite{drolsbach2024community}, and how they causally reduce the spread of misleading posts \cite{chuai2026community, slaughter2025community}.

However, Community Notes suffers from notable limitations: biased raters can manipulate the system \cite{truong2025community}, notes tend to appear on easier-to-fact-check topics, systematically biasing platform coverage \cite{wack2026laziness}, and while notes appear faster than traditional professional fact-checks, reducing their latency further remains one of the greatest levers for reducing misinformation spread \cite{slaughter2025community}. To potentially address these limitations, X’s Community Notes system has expanded note writing to AI bots while still relying on human ratings as the final arbiter of which notes are helpful and should be shown on the platform.

In this study, we analyze Collaborative Notes, a new approach to AI-assisted fact-checking introduced by X in February 2026. We specifically discuss the potential for collaborative notes to address some of the known limitations of Community Notes, such as the overall rate at which they are rated helpful \cite{de2025supernotes}, their time to consensus \cite{slaughter2025community} and their performance across topics \cite{wack2026laziness, bouchaud2025algorithmic}.

\begin{figure*}[!t]
    \centering
    \includegraphics[width=0.93\textwidth]{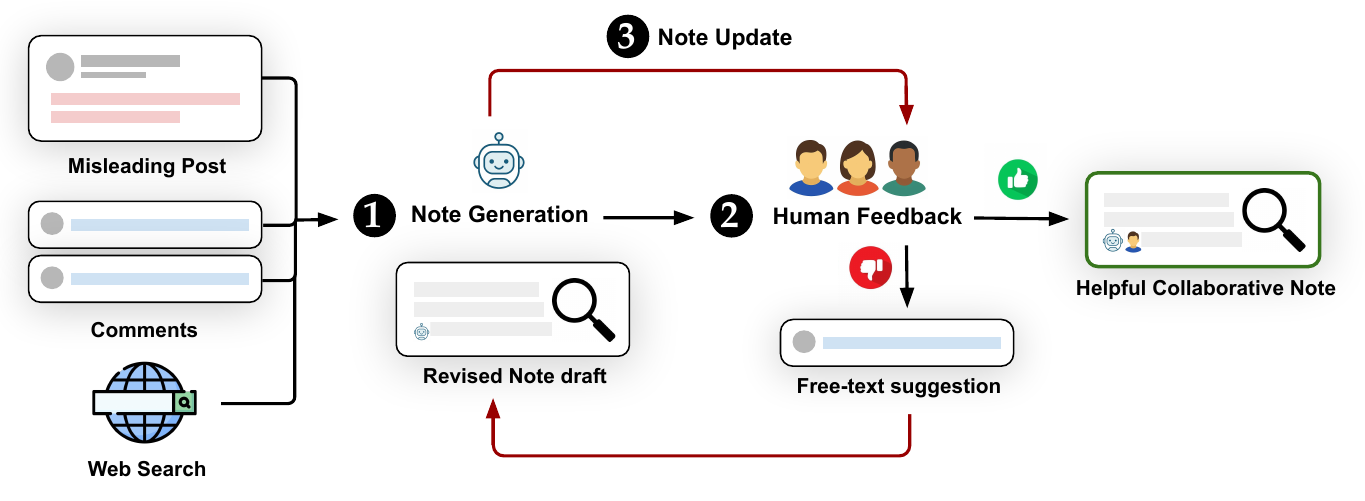}
    \caption{Lifecycle of a Collaborative Note on X. \ding{182} When a potentially misleading post is flagged, an LLM generates a draft note which is then rated by human contributors. \ding{183} Raters assess helpfulness and may optionally submit revision suggestions; a matrix factorization model scores the note based on these ratings and assigns it a status of \textit{Currently Rated Helpful}, \textit{Needs More Ratings}, or \textit{Currently Rated Not Helpful}. \ding{184} A not-helpful verdict triggers an automated revision cycle that incorporates rater suggestions and re-evaluates the note. This process repeats until the note reaches a stable helpful status or locks after two weeks. More details about the note generation, human feedback, and note update step are provided in Section~\ref{sec:background} below.}
    \label{fig:overview}
\end{figure*} 

\subsection{Human-AI collaboration in Fact-checking}

Prior work on AI-assisted fact-checking spans a wide spectrum of human involvement. At one end, \citet{de2025supernotes} and \citet{zhang2025commenotes} propose using AI to summarize human-written fact-checking content into more broadly appealing notes. At the other end, several studies propose fully automated fact-checking pipelines \cite{quelle2024perils, caramancion2023news, choi2024factgpt, kuznetsova2023generative, kuznetsova2025factchecking, li2025rag}, with \citet{zhou2024correcting} providing a Community Notes-specific instantiation and \citet{mei2026grok} investigating how users organically use Grok as a truth arbiter. Since the introduction of AI note-writers on Community Notes, early evidence suggests these systems can already outperform the average human note-writer \cite{li2026ai}.

Perhaps closest to our work is \citet{mohammadi2025ai} which demonstrates that, in a lab setting, argumentative feedback from AI can be most effective in improving notes written by humans. Similarly, \citet{juncosa2026benefit} show that small collaborative teams of human note writers can produce better fact-checking notes than individual writers. While these lab studies provide promising evidence to support collaborative notes, our study contributes to this literature by presenting an empirical study on a real-world deployment of a collaborative human-AI fact-checking system.

\section{Background}
\label{sec:background}

We begin with a brief introduction to the system, focusing only on key details to keep the description concise. The operation of Collaborative Notes on X can be abstracted as a three-stage pipeline in which AI generation, community rating, and automated scoring work together to assess and label potentially misleading posts (Figure~\ref{fig:overview}). We describe each stage below: 

\begin{itemize}
  \item[\ding{182}] \textbf{Note Generation:} When a potentially misleading post is flagged by a top note writer (a user who written enough helpful notes to have a writing impact score of at least 10) a note generation process is triggered. The system fetches the post, comments, any associated media, and existing notes. An LLM generates a note draft by evaluating the post's accuracy, validating sources, and verifying media factuality. Quality gates then reject notes that are non-English, contain spam, are stylistically unusual, or score low on an LLM-estimated accuracy signal. Failed notes are retried up to ten times before the cycle aborts.
  \item[\ding{183}] \textbf{Human Feedback:} Once published, community contributors rate the note as ``helpful,'' ``somewhat helpful,'' or ``not helpful,'' and may submit free-text suggestions for revision. A minimum of five ratings is required before the scoring pipeline can run. The scoring pipeline applies a matrix factorization to compute a per-note helpfulness score~\cite{wojcik2022birdwatch} determining whether the note reaches \textit{Currently Rated Helpful}, \textit{Needs More Ratings}, or \textit{Currently Rated Not Helpful} status.
  \item[\ding{184}] \textbf{Note Update:} The frequency of note updates depends on the status of the current version of the note. A \textit{Currently Rated Helpful} note requires a very high bar for an update (unless it is found to be inaccurate), while a \textit{Currently Rated Not Helpful} note is replaced more readily. When deciding whether to generate a new version of a note, the system re-evaluates whether the underlying story has changed, whether prior suggestions should be incorporated, and whether the proposed revision differs meaningfully from the existing note, suppressing updates that are only minor or redundant. Contributors whose suggestions were accepted or rejected are notified accordingly, and each accepted suggestion is treated as a new note with a unique identifier within the system. A note's status is locked after two weeks without any changes.
\end{itemize}

\section{Methods}
\label{sec:methods}

We collect Community Notes data from X, reflecting the full record of proposed notes and anonymized user ratings/suggestions on them as on May 5, 2026. We use qualitative and quantitative methods to analyze this dataset and answer our research questions. Below, we discuss and justify all major methodological details and choices we made. 

\subsection{Taxonomy Development and Annotation (RQ1)}
\label{methods:llm_annotations}
To characterize the content and quality of human-written suggestions on collaborative note suggestions, we develop a taxonomy. Two authors independently annotated a random sample of 100 suggestions to derive an initial taxonomy over 3-5 inductively derived dimensions, resolving disagreements through discussion. They converged on a set of 13 low-level codes (such as ``adds context", ``critique of sources", ``low actionability", etc.) organized into three higher-level dimensions: suggestion intent (characterizations the intent of the suggester, based on the text of the suggestion), evidence level (amount of evidence provided, graded by easy of verifiability), and actionability (how concrete and pointed the suggestions are). Then, they re-annotated the sample based on this initial taxonomy and did a final round of discussion and refinement to finalize the taxonomy (Table~\ref{tab:taxonomy}). We then prompted GPT-5.5 to classify every suggestion in our dataset based on the derived taxonomy, instructing the model to assign exactly one label per dimension. We evaluated the cross-annotator reliability of the labels using Cohen's $\kappa$ on an unseen test set of 100 randomly sampled suggestions. Human--human agreement was substantial across all three dimensions ($\kappa = 0.88$, $0.98$, and $0.95$ for suggestion intent, evidence level, and actionability, respectively). The agreement between the AI and two human annotators (H1 and H2) was similarly strong for suggestion intent ($\kappa_{\text{H1}} = 0.78$, $\kappa_{\text{H2}} = 0.85$) and evidence level ($\kappa_{\text{H1}} = 0.89$, $\kappa_{\text{H2}} = 0.91$), and moderate for actionability ($\kappa_{\text{H1}} = 0.64$, $\kappa_{\text{H2}} = 0.61$). All final prompts are included in Appendix \ref{app:llm-prompts}.

To determine whether each suggestion was incorporated into the revised note, we use a prompt from the Community Notes collaborative note generator implementation, which takes the previous note, revised note, and suggestion as input and outputs one of three labels: \textit{yes}, \textit{partially}, or \textit{no}. We run this using Grok-4.20-reasoning. A suggestion is labeled \textit{yes} or \textit{partially} only if the revised note differs from the previous one in a way that reflects the specific suggestion; when in doubt, the model defaults to \textit{no}. For reproducibility, the exact prompt we used is included in Appendix \ref{app:grok_prompt}.

\subsection{Measuring Performance Metrics (RQ2, RQ3)}
We consider two key metrics when analyzing the performance of notes on Community Notes. First, we calculate the note helpfulness, a note is only shown publicly alongside a misleading post if it is judged to be helpful. Second, we also consider the time it takes for a note to reach a status (helpful or not), as research shows that this time has a strong impact on the potential of these notes to reduce the spread and engagement with misleading content on the platform \cite{slaughter2025community}.

\subsubsection{Note Helpfulness:} To measure how helpful a note is rated by Community Notes contributors, we compute a \textit{note intercept} for each note (all versions of collaborative notes and non-collaborative notes) using the Core Matrix Factorization scorer in the pre-scoring phase of the Community Notes algorithm~\cite{wojcik2022birdwatch}.\footnote{\url{https://communitynotes.x.com/guide/en/under-the-hood/ranking-notes\#complete-algorithm-steps}} We run the algorithm on the entire historical Community Notes dataset as on May 05, 2026, using the official implementation of the algorithm (GitHub commit \texttt{03cbdf9})\footnote{https://github.com/twitter/communitynotes/}, open-sourced under an Apache 2.0 license. Each rating (``not helpful", ``somewhat helpful", and ``helpful") is modeled as a combination of a global baseline, a per-note intercept, and rater-specific factors that capture individual rating tendencies. The matrix factorization yields an estimate of the note intercept for each note; a higher value indicates that the note is rated as helpful broadly across raters with diverse viewpoints. We refer to this note intercept as the helpfulness score throughout our findings. We restrict the algorithm to the pre-scoring phase because it reduces runtime to under 24 hours on the full dataset, allowing us to include more recent data in our analysis, and the outputs are comparable with outputs from earlier algorithm versions used in prior literature \cite{wojcik2022birdwatch} which only had a single phase.

\subsubsection{Time to Consensus:} 
We use the publicly-released status history of the notes to measure the elapsed time from creation to the first transition out of \textit{Needs More Ratings} (NMR) status, either to \textit{Currently Rated Helpful} (CRH) or \textit{Currently Rated Not Helpful} (CRNH). We only consider notes that have reached CRN or CRNH status. Collaborative notes require a different treatment because multiple note versions form a revision-chain on the same tweet. Therefore, we measure time elapsed from the creation time of the first version of a collaborative note to the first transition out of the  \textit{Needs More Ratings} (NMR) status over any subsequent revision of that note. Similar to non-collaborative notes, if no version of a collaborative note on a post reaches consensus status (CRH or CRNH), we exclude all collaborative note versions on that post from our analyses.

When comparing collaborative notes against human- or AI-written notes, we restrict the comparison to notes published after the date the first collaborative note appeared, ensuring a temporally comparable sample.

\subsection{Regression Analysis (RQ2)}

To understand what kinds of suggestions and which raters are most impactful in improving collaborative notes over consecutive versions, we perform regression analyses at the version-to-version transition level. 

We fit two OLS regressions with cluster-robust standard errors clustered at the post level to account for multiple transitions per post. The outcome in both models is the change in note helpfulness score (intercept) between consecutive versions, restricted to transitions where at least one suggestion was incorporated (see Section \ref{methods:llm_annotations}) and the intercept was computed for both notes ($n=285$).

First, we regress the change in helpfulness scores (intercept) on the content features in the text of the incorporation suggestions. We encode each suggestion intent category (Challenge Claim, Add Context, Source Critique, Meta Policy, Formatting Edit, AI Resistance, Value Judgment) as a binary indicator equal to 1 if any incorporated suggestion in the transition carried that intent label. Similarly, we collapse actionability and verifiability into binary variables: actionability is encoded as 1 if any suggestion was rated as having high or medium actionability, and verifiability is encoded as 1 if any suggestion included links or was otherwise verifiable. We control for the note's baseline helpfulness score (intercept) and the total number of suggestions in the transition. The baseline score is included to account for ceiling effects, as notes with higher initial scores have less room to improve. The total number of suggestions is included to isolate the effect of which types of suggestions were incorporated rather than how many. We remove transition categories that have occurred fewer than 20 times, for brevity.

Next, we regress the change in helpfulness scores on rater characteristics. Each predictor is a binary indicator equal to 1 if any suggester in the transition exceeded a threshold: above-median tenure (days since first rating), above-median rating volume (total ratings cast on collaborative notes), ideologically extreme lean ($|\text{rater factor}| > 0.5$), and writing ability (enrollment state = earnedIn). Control covariates are identical to the previous model.

\subsection{Topic Annotation (RQ3)}
Finally, to understand topical differences in the performance of Collaborative Notes, we annotated the topics of the rater provided suggestions. Following the approach adopted by \citet{bouchaud2025algorithmic}, topic annotations for notes were generated using \texttt{cardiffnlp/tweet-topic-latest-multi} \cite{antypas2022twitter}, a multi-label tweet topic classifier that produces a probability that the note belongs to each of 19 topic categories, such as News \& Social Concern, Sports, Music and Fitness \& Health. While trained on tweet data, we find that this model performs well on note data as well given similar length constraints and present a validation of our results in Appendix \ref{appendix:topics} following the approach in \citet{bouchaud2025algorithmic}. Notes not already in English were first machine-translated using \texttt{facebook/m2m100\_418M} \cite{fan2021beyond}, a multilingual neural translation model, with the translated text passed to the topic classifier. The primary topic is defined as the highest-scoring category from the classifier.

\section{Results}\label{sec:results}
Next, we present findings related to our three research questions. First, we characterize the kinds of suggestions provided by raters on collaborative notes and who they come from. Second, we trace how collaborative notes change across versions and how their evolution is shaped by chain length, suggestion type, and rater characteristics. Finally, we compare the performance of collaborative notes against human- and AI-written baselines.

\subsection{Characterizing Human Input (RQ1)}
\label{sec:results_human_input}

\begin{figure}[htp]
    \centering
    \begin{subfigure}{\columnwidth}
        \centering
        \includegraphics[width=\columnwidth]{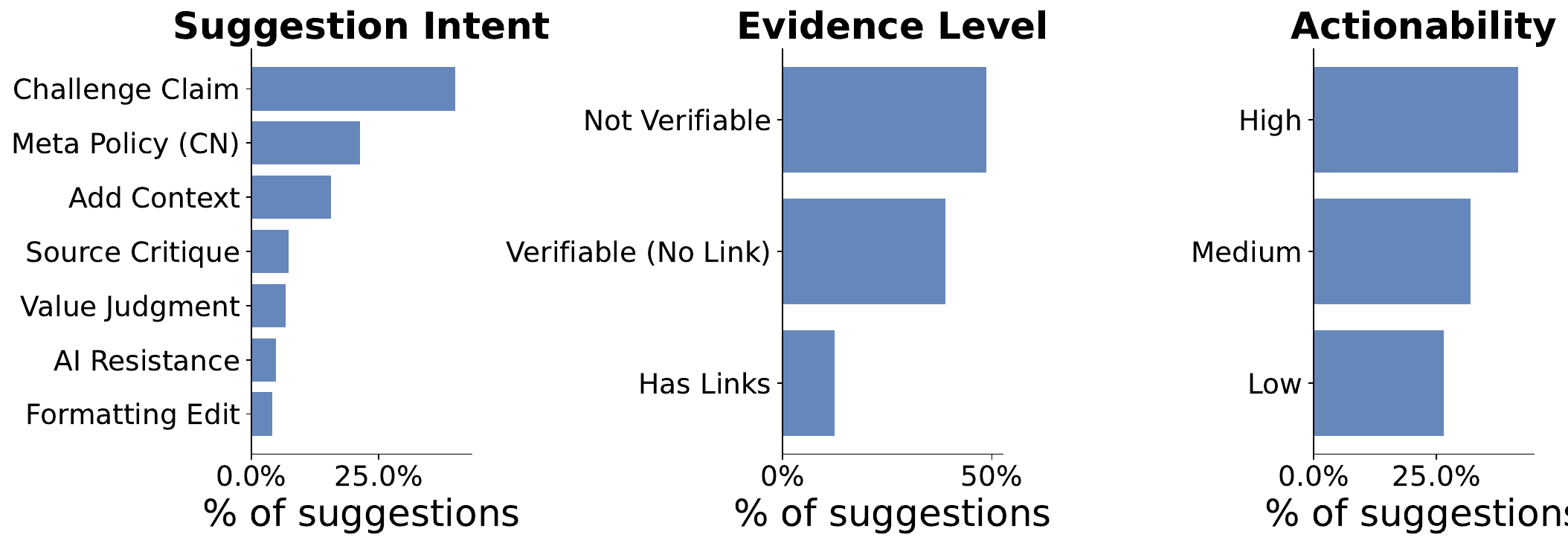}
        \caption{Fraction of suggestions containing each intent, evidence level, and level of actionability (n=10,458).}
        \label{fig:taxonomy_prevalence}
    \end{subfigure}
    
    \vspace{1em}
    
    \begin{subfigure}{\columnwidth}
        \centering
        \includegraphics[width=\columnwidth]{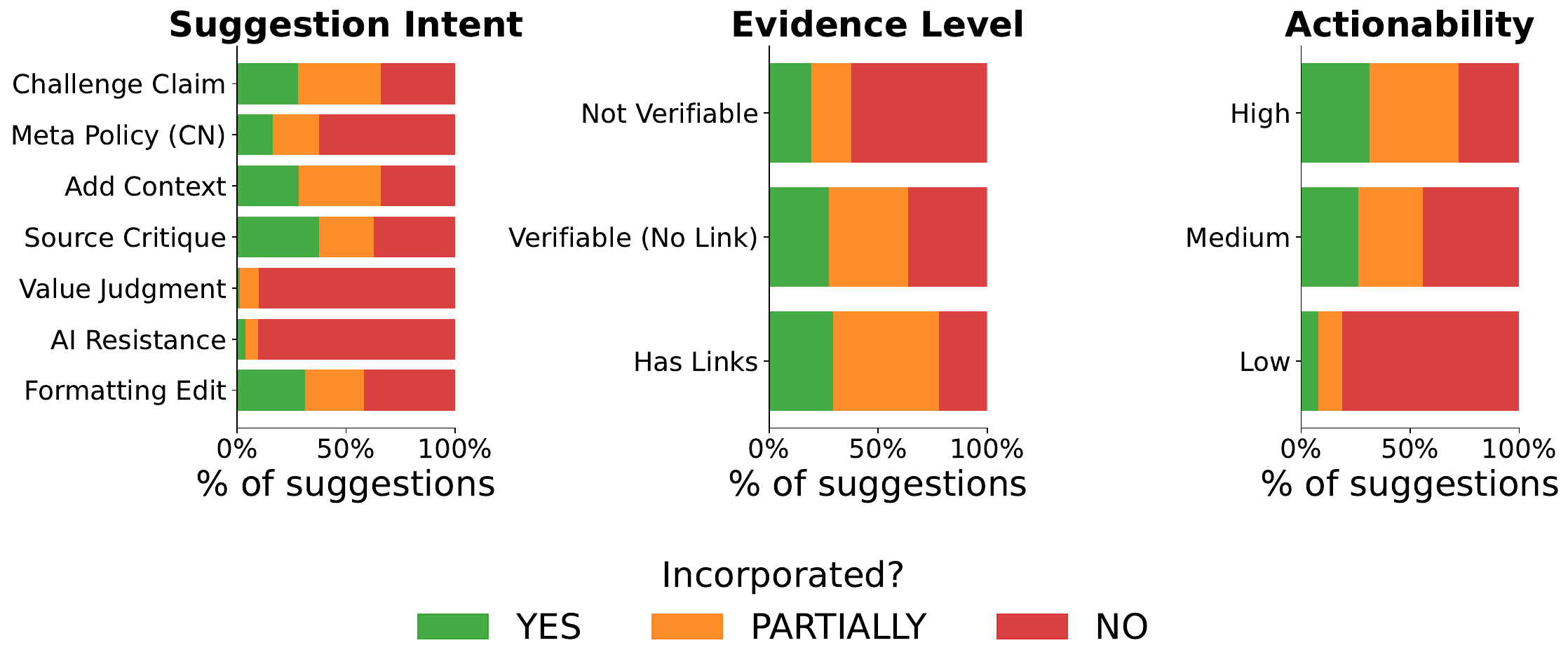}
        \caption{Fraction of suggestions with a given intent, evidence level, or level of actionability that are incorporated into the collaborative note (n=7,004).}
        \label{fig:incorporation_by_taxonomy}
    \end{subfigure}
    
    \caption{Frequency and success rates for suggestions with different intents, evidence levels, and levels of actionability.}
    \label{fig:taxonomy_analysis}
\end{figure}

We find that the intent of suggestions made by humans on collaborative notes varies widely, and can be divided into 7 broad types. As shown in Figure \ref{fig:taxonomy_prevalence}, the most common types of suggestion typically challenge claims made in the post (40.1\%) or reflect the rater's judgment on whether the note is needed or not for a given post, i.e., a meta comment about Community Notes policies (21.4\%) or request the addition of more or additional context (15.7\%). A smaller portion of suggestions critique sources used in AI notes (7.3\%), reflect moral or value-based judgments that express personal opinions (6.7\%), or provide editorial suggestions (4.1\%). Interestingly, we also note that a subset of suggestions express explicit resistance against the use of AI in Community Notes (4.8\%). For a more detailed description of these categories, including examples, refer to Table~\ref{tab:taxonomy}. Notably, almost a majority of suggestions (48.8\%) are not independently verifiable, they neither include supporting links nor constitute falsifiable claims, limiting the degree to which their validity can be assessed.

As described in Section \ref{sec:background}, not all suggestions are incorporated into the generation of a new note. Relatedly, we find that certain types of suggestions are more likely to be incorporated than others. Figure \ref{fig:incorporation_by_taxonomy} shows while suggestions that critique sources or offer editorial advice may be low in volume, they are in fact more frequently incorporated into new versions (37.7\% and 31.3\% respectively). Conversely, value-laden suggestions, or expressions of resistance against the use of AI, while useful for capturing community sentiment, are least likely to be incorporated into newer versions (1.2\% and 4\% respectively). Consistent patterns across evidence levels and actionability also suggests that raters whose contributions are grounded in verifiable information including links and are generally more actionable play an outsized role in shaping the final form of collaborative notes. Suggestions with low levels of actionability are more than twice as likely to be rejected than highly actionable suggestions. 
\\

\begin{figure}[htp]
    \centering
    \includegraphics[width=\columnwidth]{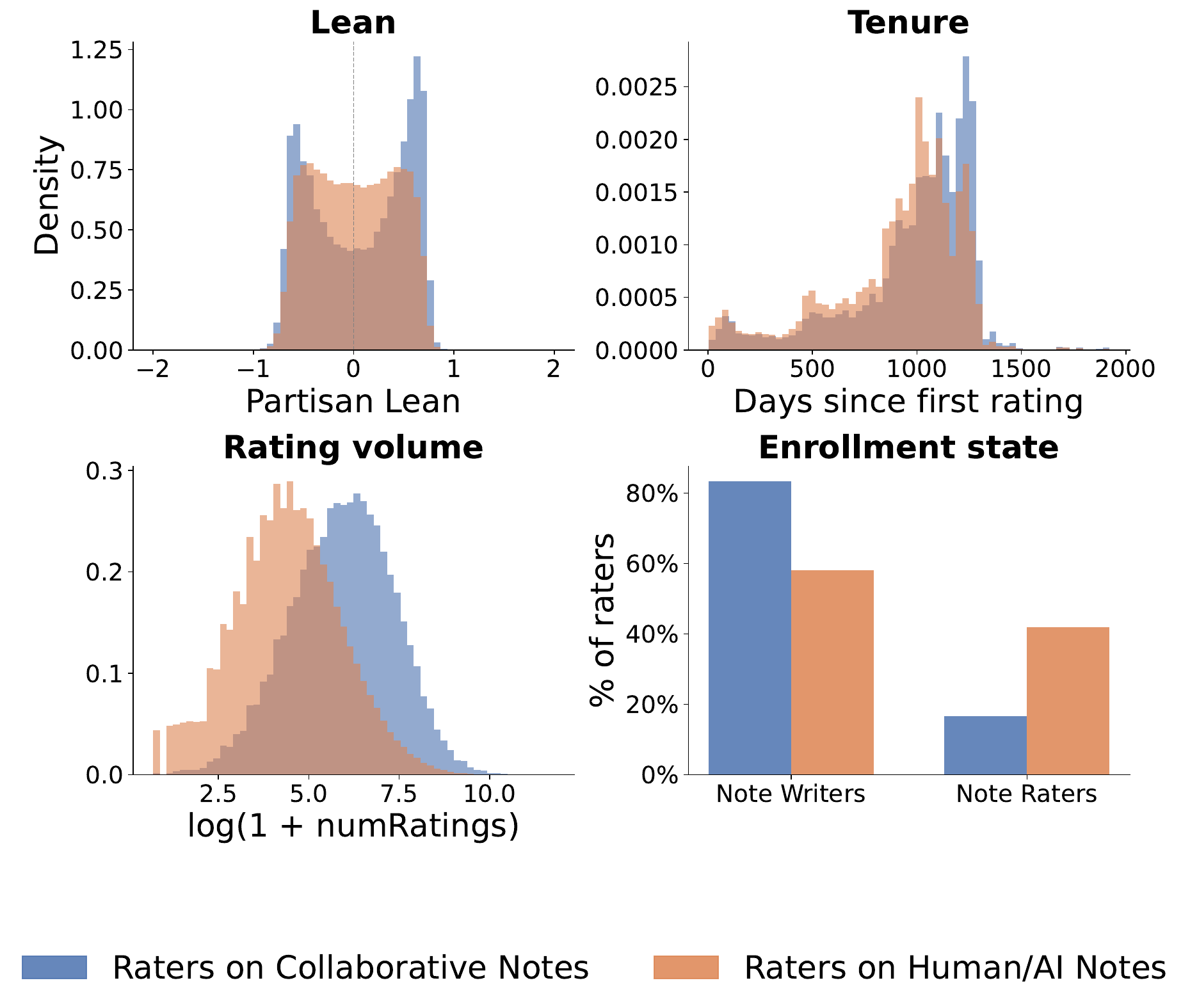}
    \caption{Characteristics of users who rated collaborative vs. non-collaborative notes. \textit{Partisan lean} indicates a user's latent ideology score as calculated by the Community Notes algorithm; \textit{tenure} indicates number of days since the user first provided a rating; \textit{volume} indicates total number of ratings made by the user; \textit{enrollment state} indicates whether a user has successfully rated enough notes to earn the ability to write notes.}
    \label{fig:rater_demographics}
\end{figure}

Having characterized the content of the suggestions, we now turn to the contributors who provide such feedback on collaborative notes. We find that these contributors represent a markedly different population from those who rate non-collaborative human- or AI-written notes over the same period. As shown in Figure \ref{fig:rater_demographics}, raters on collaborative notes tend to exhibit more extreme rater factors ($b_{\text{collab}}~=~0.664, b_{\text{non-collab}}~=~0.551$, where $b$ is the bimodality coefficient, which measures the extent to which a distribution is bimodal, i.e., has two peaks rather than one)---indicative of a more polarized political lean based on prior rating behavior---and have longer platform tenure ($\mu_{\text{collab}}~=~983.8 \pm 2.0, \mu_{\text{non-collab}}~=~908.1 \pm 0.7$) and substantially higher overall rating volumes ($\mu_{\text{collab}}~=~936.2 \pm 13.4, \mu_{\text{non-collab}}~=~252 \pm 1.7$). Furthermore, the majority of these raters have already earned the ability to write notes themselves (83.3\% $\pm 0.3\%$), suggesting that contributions to the collaborative note system are predominantly driven by experienced and polarized ``veterans'' who were already active on the platform prior to the system's introduction. Furthermore, as \citet{nudo2026hyperactive} points out, contributions from such hyperactive minorities can often result in noisy and unstable predictions of overall note helpfulness. All intervals are bootstrapped 95\% CIs.

\subsection{Characterizing Note Evolution (RQ2)}
\label{sec:results_note_evolve}
\begin{figure}[t]
    \centering
    
    \begin{subfigure}{0.95\columnwidth}
        \centering
        \includegraphics[width=\columnwidth]{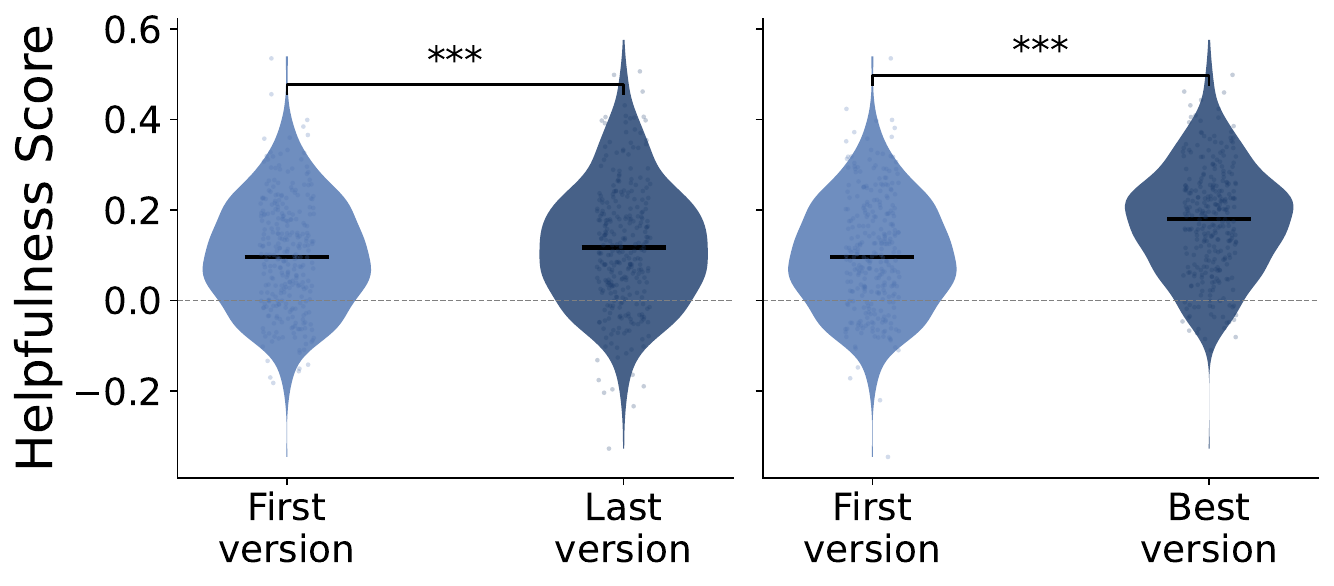}
        \caption{Mean Improvement in note helpfulness score relative to first version. \textit{Last} version indicates final collaborative note drafted by AI; \textit{Best} version indicates Collaborative Note achieving the highest helpfulness score across all iterations. Wilcoxon signed-rank tests reveals that both improvements are statistically significant ($p<0.001$).}
        \label{fig:best_last}
    \end{subfigure}

     \vspace{1em}

    \begin{subfigure}{0.95\columnwidth}
        \centering
        \includegraphics[width=\columnwidth]{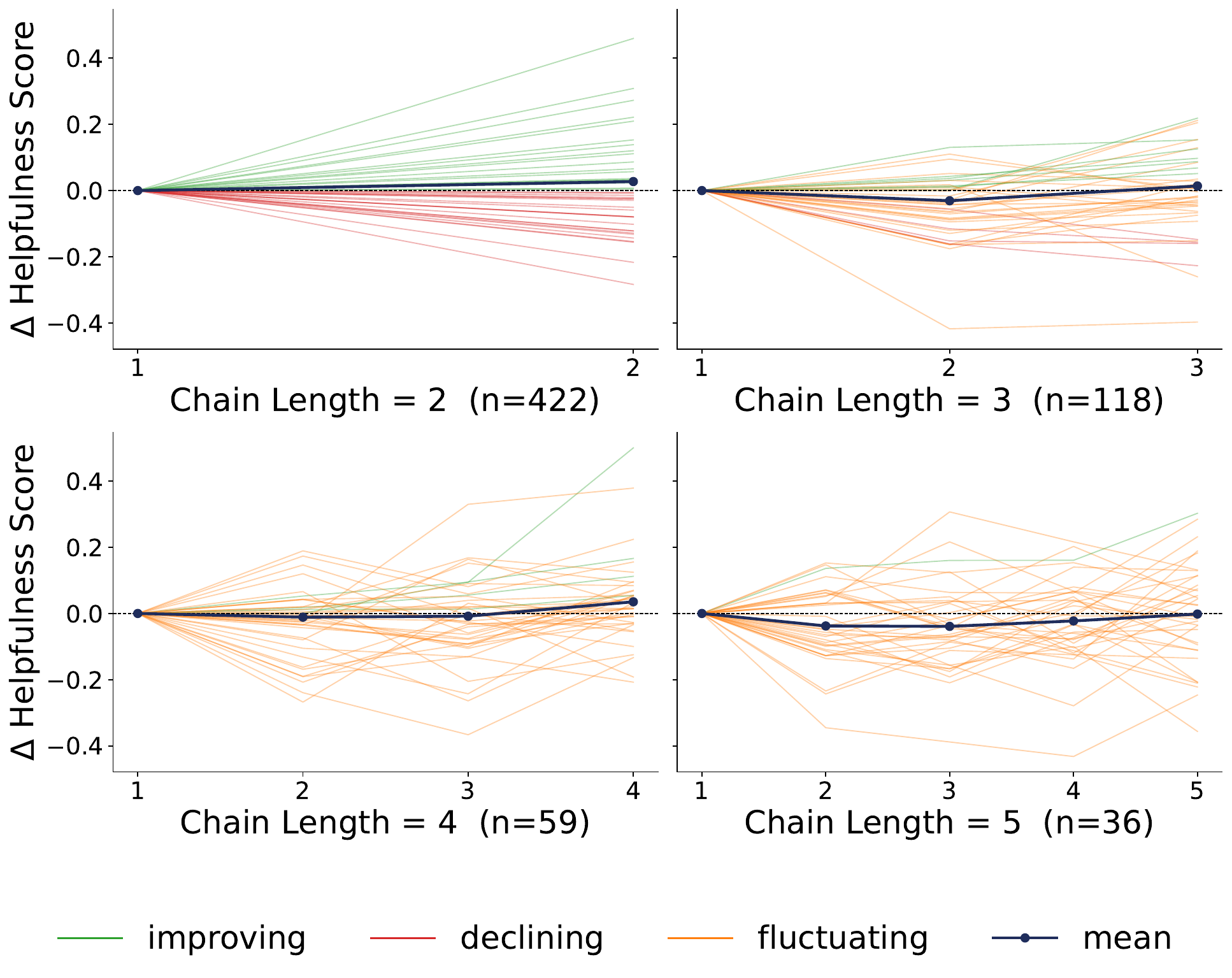}
        \caption{Changes in helpfulness scores for collaborative notes of differing chain lengths. Colored lines represents individual notes. Green indicates a monotonic increase, red indicates a monotonic decrease, and orange indicates non-monotonic change.}
        \label{fig:note_trajectories}
    \end{subfigure}
    
    \caption{Evolution of collaborative notes across versions.}
\end{figure}

Collaborative notes are iteratively revised as new ratings and rater suggestions become available, resulting in a sequence of versions that reflects the cumulative effect of human input. Figure \ref{fig:best_last} shows that both the best and final versions of a collaborative note are rated as significantly more helpful than the initial AI-drafted version ($p<0.001$), providing evidence that human feedback drives meaningful quality gains. However, as Figure \ref{fig:note_trajectories} illustrates, this improvement is rarely monotonic---individual note trajectories frequently fluctuate across versions before converging on a higher quality outcome. We investigate three factors that may explain variation in how notes evolve: chain length, suggestion type, and rater characteristics.

\begin{figure}[h!]
    \centering
    \includegraphics[width=0.9\columnwidth]{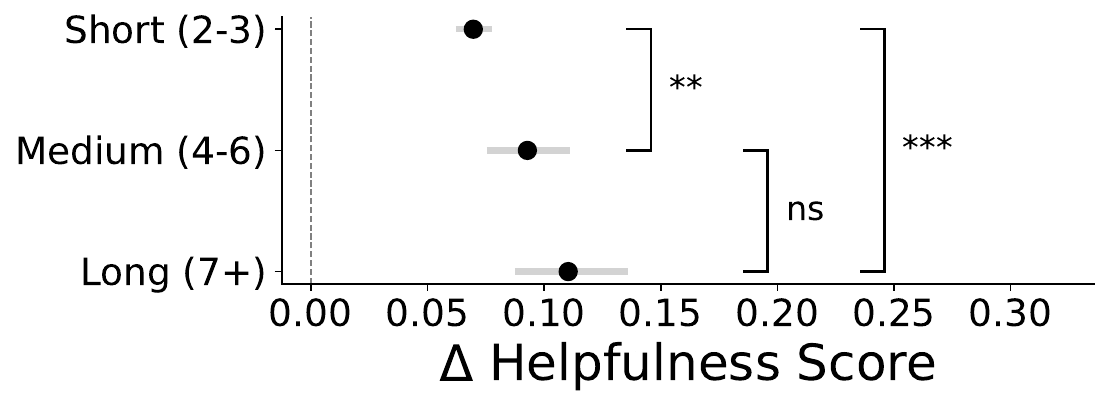}
    \caption{Mean Improvement in Note Helpfulness Score relative to first version, by chain length. Collaborative notes with longer revision histories show significantly greater quality improvement over their lifetime. Error bars represent bootstrapped 95\% confidence intervals.}
    \label{fig:chain_length}
\end{figure}

\textbf{Chain Length.} The number of revisions a note undergoes is positively associated with its overall quality improvement. As shown in Figure \ref{fig:chain_length}, mean intercept change increases monotonically across chain length buckets, and pairwise comparisons confirm that these differences are statistically significant (Mann-Whitney U, BH-corrected): short chains (2--3 versions) differ significantly from medium chains (4--6 versions; $p < 0.01$) and from long chains (7+ versions; $p < 0.001$). This pattern suggests that sustained revision, rather than any single update, is the primary driver of note quality gains.

\begin{figure}[h!]
    \centering
    \includegraphics[width=\columnwidth]{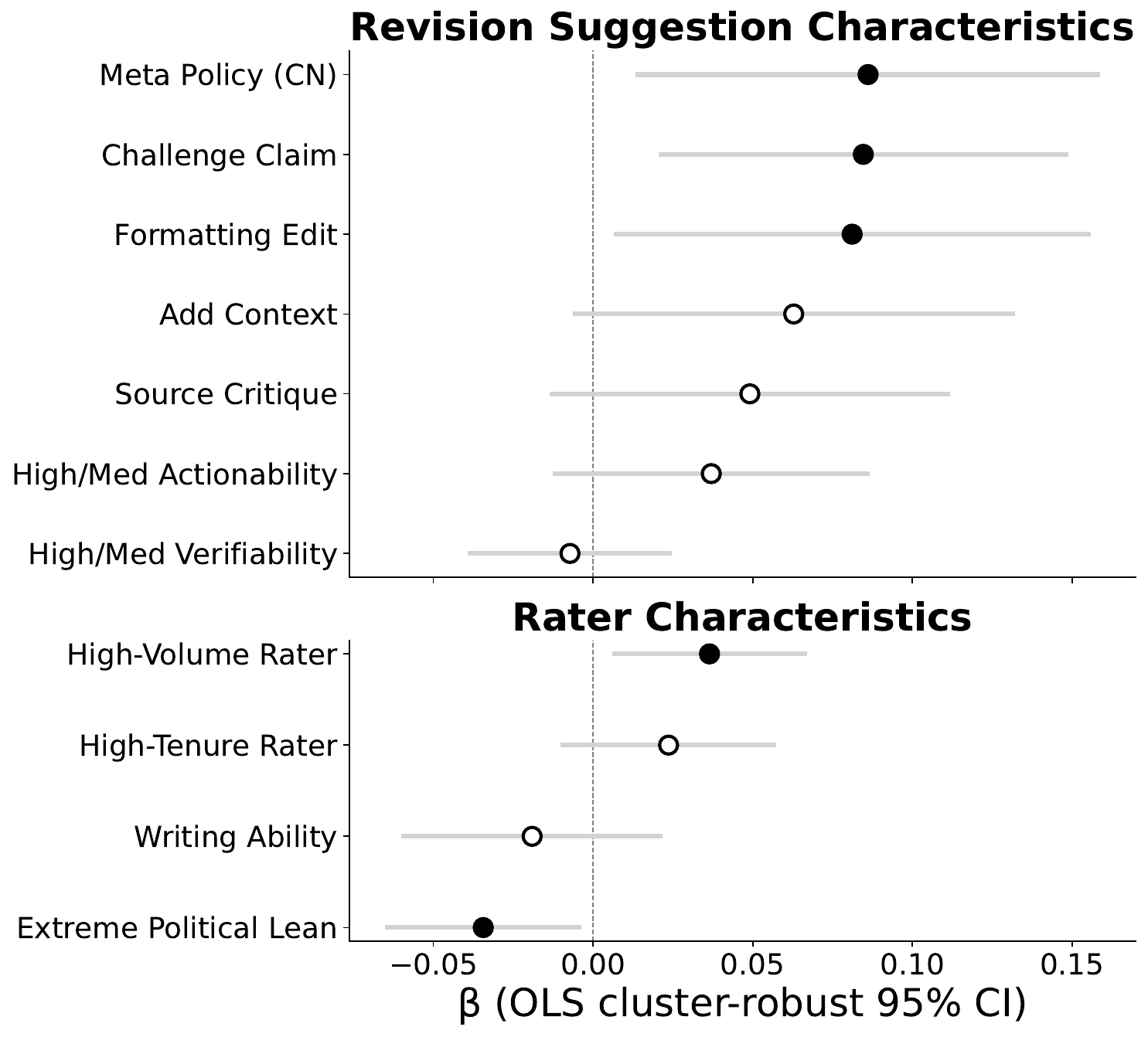}
    \caption{Associations between note improvement in an iteration and (top) content of rater's suggestion, (bottom) characteristics of the rater. $\beta$ refers to estimate from linear regression of improvement on indicator for whether suggestion included feature or rater held characteristic, controlling for total number of suggestions made. Shows all transitions where suggestion was incorporated ($N = 285$), and both pre- and post-suggestion note were scored with the Community Notes algorithm.}
    \label{fig:suggestion_types}
\end{figure}

\begin{figure}[t]
    \centering
    \includegraphics[width=\columnwidth]{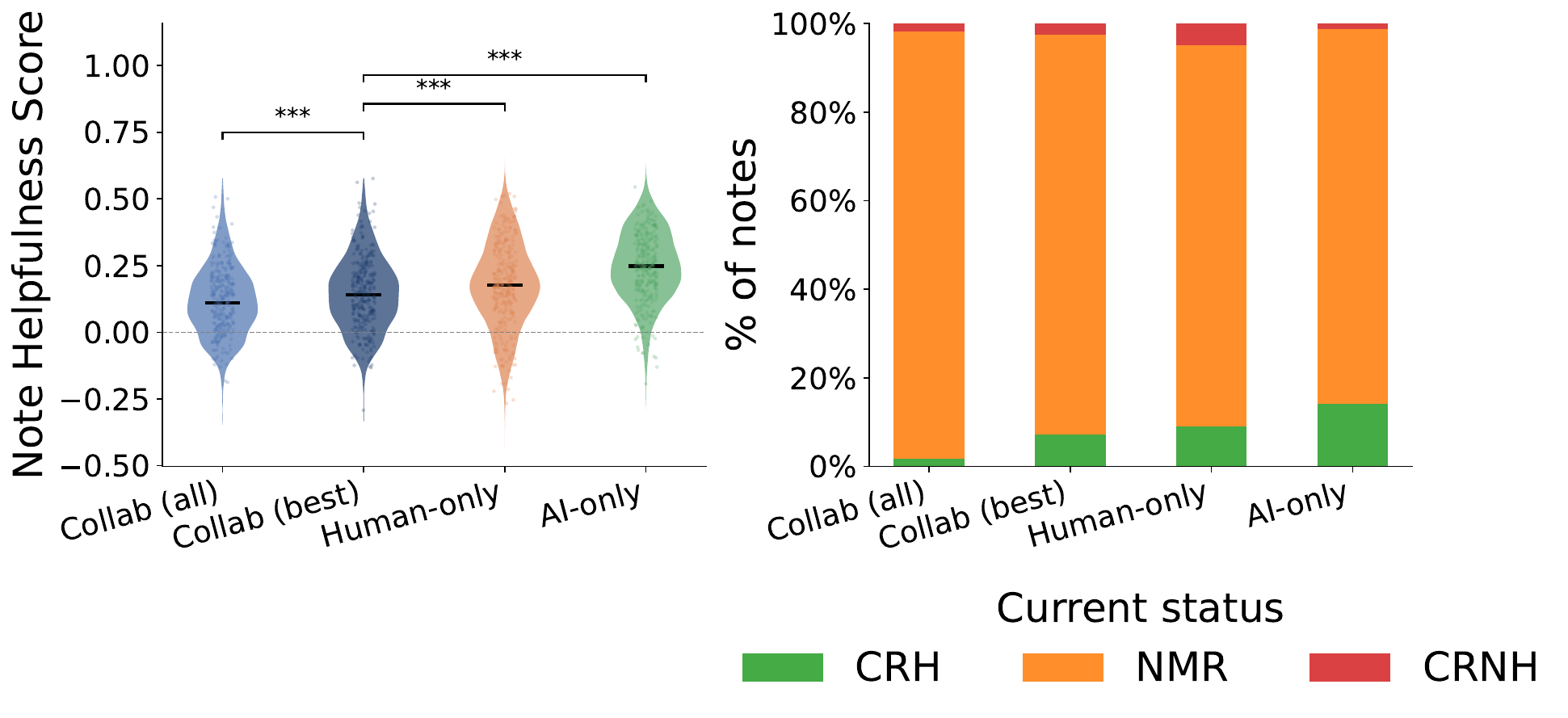}
    \caption{Performance of collaborative notes relative to non-collaborative notes. \textit{Collab (all)} refers to all versions of all collaborative notes; \textit{Collab (best)} refers to the best version of each collaborative note. Left plot shows distribution of note helpfulness score, while right shows percent of notes labeled as \textit{Currently Rated Helpful} (CRH), \textit{Needs More Ratings} (NMR), or \textit{Currently Rated Not Helpful} (CRNH).  Collaborative notes have significantly lower intercepts than non-collaborative notes. Mann-Whitney U tests (BH-corrected) reveal that pairwise differences in intercept are statistically significant ($p<0.001$).}

    \label{fig:collab_vs_others}
\end{figure}

\textbf{Suggestion Type.} Three intent categories are associated with statistically significant improvements: Challenge Claim ($\beta  = +0.085, p < .01$), Meta Policy (CN) ($\beta = +0.086, p < .05$), and Formatting Edit ($\beta = +0.081, p < .05$). Add Context and Source Critique show positive but non-significant estimates. Neither actionability level nor verifiability independently predicts improvement, suggesting that what the suggestion says matters more than whether it includes a source link. A strong negative coefficient on baseline intercept ($\beta = -0.67, p < .001$) implies that lower-quality notes have more room to improve. 

\textbf{Rater Characteristics.} Finally, we examine whether the characteristics of raters providing suggestions are associated with their impact on note helpfulness. High-volume raters, i.e., those with above-median rating activity, are associated with larger improvements ($\beta = +0.037, p < .05$), consistent with more experienced contributors offering better feedback. Conversely, suggestions from ideologically extreme raters ($|\text{lean}| > 0.5$) predict smaller improvements ($\beta = -0.034, p < .05$), suggesting that highly partisan perspectives may steer notes in directions that are less broadly helpful, consistent with \citet{nudo2026hyperactive}. Tenure and the ability to write notes are not independently significant after controlling for baseline quality and suggestion volume. This points to the potential of newer contributors, who may not have long tenure or the ability to write notes, to have a positive impact on the system through suggestions.

\subsection{Effectiveness of Collaborative Notes (RQ3)}
\label{sec:results_compare}

\begin{figure*}[t!]
    \centering
    \begin{subfigure}[t]{0.45\textwidth}
        \centering
        \includegraphics[width=\columnwidth]{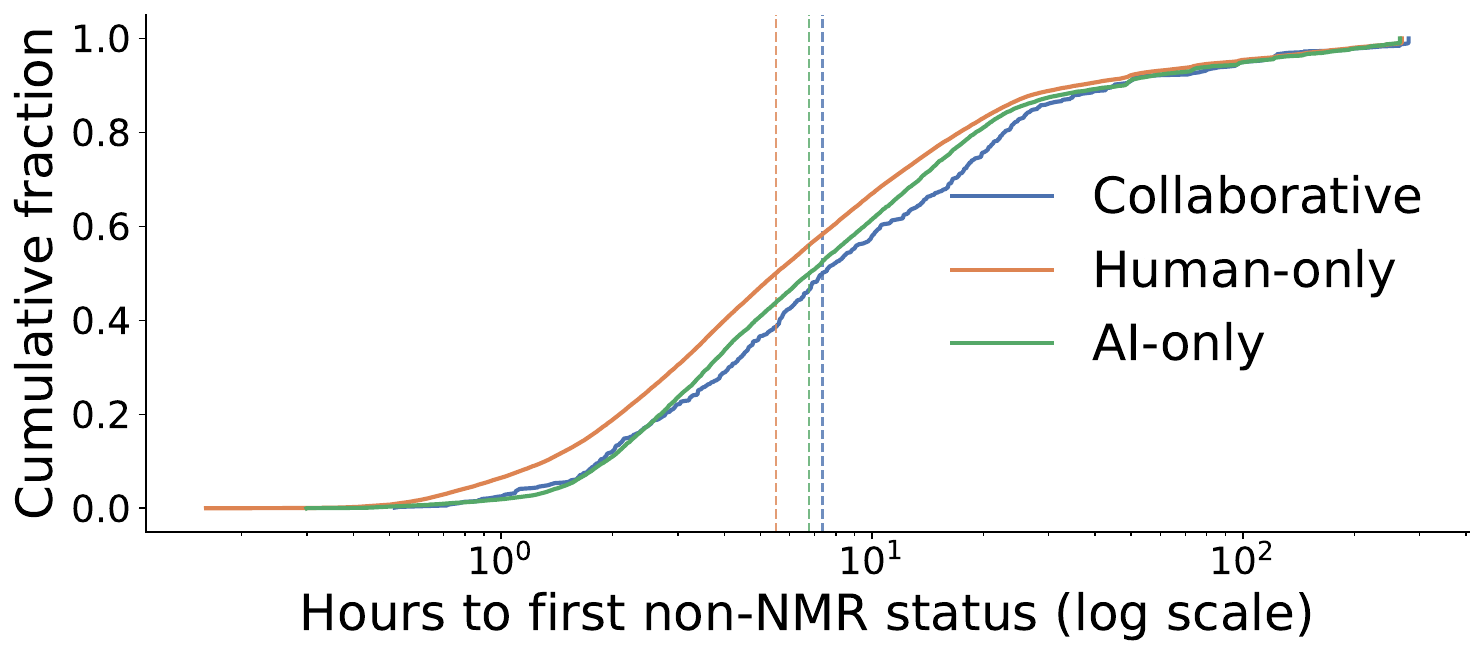}
    \caption{Cumulative distribution of time needed for notes to reach a\textit{ Currently Rated Helpful} or \textit{Currently Rated Not Helpful} status.}
    \label{fig:time_to_consensus}
    \end{subfigure}
    \qquad \qquad
    \begin{subfigure}[t]{0.46\textwidth}
        \centering
        \includegraphics[width=\columnwidth]{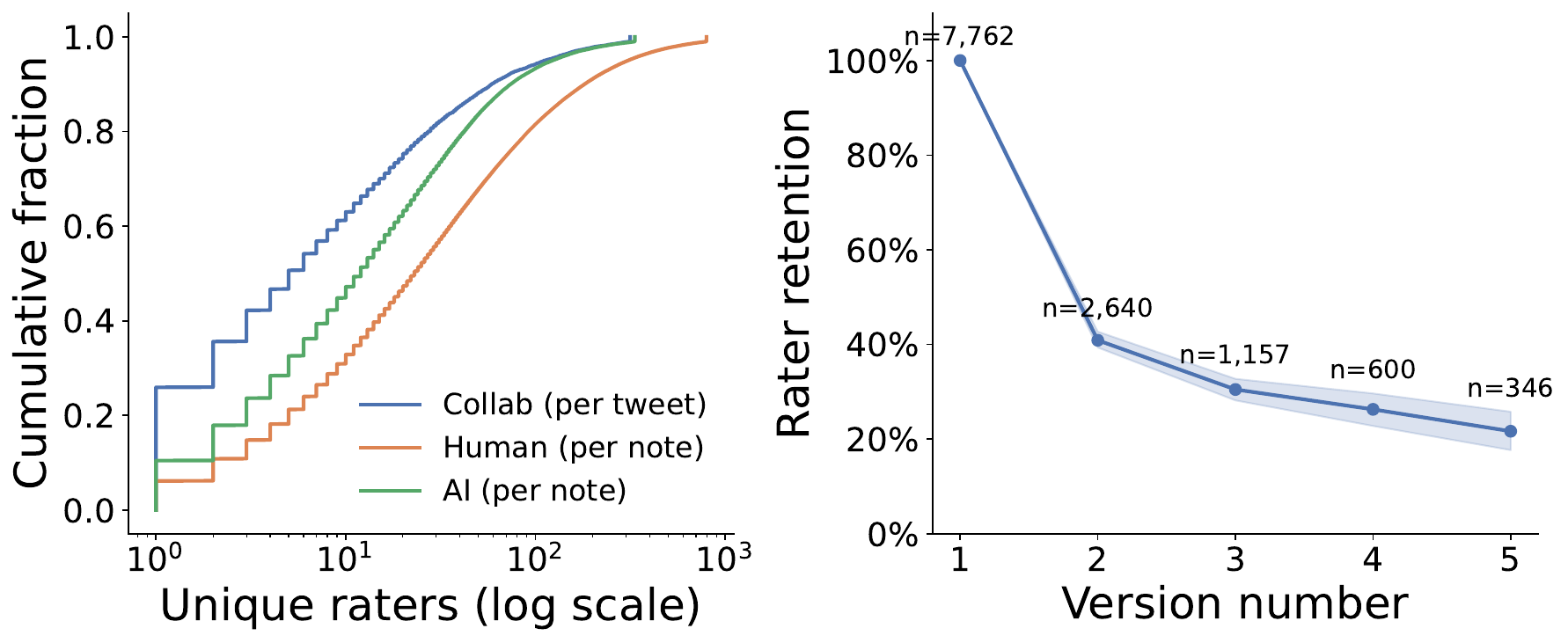}
        \caption{Left shows cumulative distribution of unique raters; right shows average number of raters who rated both the first version of a note and a given version number.}
        \label{fig:rater_reach}
    \end{subfigure}
    \caption{Speed of rating accumulation for different note types, number of raters for different note types, and rater retention for collaborative notes.}
    \label{fig:reasons}
\end{figure*}

Despite the iterative refinement process described in the previous section, collaborative notes do not, on average, match the overall helpfulness of human- or AI-written notes. As shown in Figure \ref{fig:collab_vs_others}, collaborative notes exhibit the lowest average note intercepts across all three note types, suggesting that the collaborative process, in its current form, does not yet close the gap with non-collaborative baselines. Only 1.8\% of collaborative notes achieve a helpful status, compared to 14.1\% for AI-written notes and 9.0\% for Human-written notes. Additionally, the overall under-performance of collaborative notes is uniform across all topic domains, as shown in Figure \ref{fig:collab_vs_others_topics}. While the gap in performance between collaborative notes and human- and AI-written notes is relatively smaller on topics such as fitness, science, and technology, the consistent underperformance of collaborative notes suggests that this trend is indeed topic-agnostic. Next, we present a few possible explanations for this lack of performance.

\subsubsection{Longer time to consensus.} Collaborative notes face challenges in reaching rater consensus in a timely manner. Figure \ref{fig:time_to_consensus} shows that collaborative notes that do reach a non-NMR status take longer than non-collaborative notes to do so, indicating that the iterative revision process introduces delays that may reduce the practical utility of these notes in time-sensitive fact-checking contexts.

\subsubsection{Lack of ratings.} In Figure \ref{fig:rater_reach} we find that collaborative notes attract significantly fewer raters ($\mu = 25.7$, median $=5$) than human ($\mu = 75.6$, median $=23$) and AI-written notes ($\mu = 33.1$, median $=12$), even when counting unique raters across all versions of a collaborative note. We speculate that this may result from a combination of factors, ranging from a general aversion to collaborating on AI-generated notes (as evidenced by the ``AI Resistance" category in our taxonomy) as well as greater burdens associated with the expectation to provide free-text suggestions on such notes. We also note that rater engagement drops sharply across successive versions of a collaborative note. Among tweets that received at least two versions of a collaborative note, only 40.9\% (95\% CI: [39.4\%, 42.7\%]) of raters who rated the first version returned to rate the second, falling to 30.5\% by version 3 and 21.7\% by version 5. This suggests that the pool of engaged raters is largely non-overlapping across versions: the majority of raters who evaluate an early draft do not revisit the note after it is updated, leaving later versions to be assessed by a different audience.

\subsubsection{Complementary coverage.} While these limitations may suggest that collaborative notes are a weaker substitute for non-collaborative ones, we find evidence that they play a complementary role in the ecosystem, often targeting posts that do not attract human- or AI-only notes.

Collaborative notes tend to target a distinct subset of posts that have fewer non-collaborative notes. Of the 10,600 posts with at least one collaborative note, 76.5\% did not have any human- or AI-only notes proposed. Similarly, 65.9\% of posts with a publicly shown helpful collaborative note did not have any other helpful notes. Whether this reflects greater post difficulty or differences in which posts attract collaborative effort remains an open question, but the pattern suggests that collaborative notes serve a complementary function, expanding fact-checking coverage to posts that non-collaborative notes do not reach.

\section{Discussion}
\label{sec:discussion}
This work provides an empirical characterization of a human-AI collaborative fact-checking system, examining the nature of human contributions, the dynamics of note improvement over time, and the comparative effectiveness of the resulting notes. While the collaborative process engaging human contributors does drive meaningful quality improvements, collaborative notes face structural limitations that constrain its overall effectiveness.

Our analysis of human input reveals that participation in the collaborative note process is concentrated among experienced, polarized and highly active platform users. While this veteran skew may confer certain quality benefits---raters with many prior ratings tend to provide more impactful suggestions---it also raises questions about the representativeness of the feedback and the scalability of the system. The taxonomy of suggestions further highlights that not all human input is equally constructive: claim challenges, source critiques and contextual additions are the most likely to be incorporated and to improve note quality, while suggestions carrying more subjective value judgments and expressions against the use of AI are likely to not be incorporated into new versions.

The analysis of note evolution confirms that sustained revision is beneficial but that the path to improvement is rarely smooth. Individual note trajectories frequently fluctuate across versions, and the rate of improvement is sensitive to both the type of suggestions received and the demographics of the raters providing them. These findings suggest that the quality of the collaborative process is not simply a function of how many revisions occur, but of the quality and consistency of the human input at each step.

Despite these improvements, collaborative notes do not yet match the helpfulness of human- or AI-written notes, and they take longer to reach rater consensus. A key structural issue is the lack of ratings received by collaborative notes on average, relative to human or AI-written notes. Yet their importance should not be understated: collaborative notes extend coverage to posts that would otherwise go without any note, filling gaps that the existing note ecosystem leaves behind. Improvements to the collaborative note pipeline therefore carry a dual benefit: raising the quality of notes that already exist and bringing helpful, consensus-reaching notes to posts that would otherwise have none.

\subsection{Design Implications}
\label{sec:design_implications}
Our findings point to concrete opportunities for improving the system. Most contributors who rated an early version do not return to rate subsequent ones, meaning the system effectively starts from scratch with each revision. One mitigation is to treat ratings cumulatively, carrying forward signal from prior versions. A principled approach would be to use the rating shifts of contributors who rate multiple versions as anchors to calibrate how the broader rater pool would likely respond to the updated note. To make this effective, the platform could nudge raters of earlier versions via notifications to return and rate subsequent ones.

Second, raters could be guided toward providing more effective suggestions. Our results demonstrate that certain suggestion types are substantially more likely to improve note quality than others. Surfacing this information to raters at the point of contribution, for instance through lightweight prompts or structured suggestion templates, could help shift the distribution of feedback toward more actionable and impactful inputs. As proposed in previous work \cite{li2025scaling}, tools may also be developed to assist raters provide more effective ratings and suggestions, further reducing the barriers to contribution.

Third, decisions on which suggestions to incorporate or discard may be grounded in historical effectiveness of such suggestions, as we demonstrate in Section \ref{sec:results_human_input}, instead of relying entirely on an LLM-judge. This may also be aided by the use of predictive models such as those used in \citet{de2025supernotes}.

Finally, all efforts to broaden participation of raters in the collaborative notes program is likely to drive gains. In particular, reducing frictions by explicitly lowering expectations to provide free-text suggestions, or by nudging newer raters to engage in collaboration might not only increase the overall ratings on collaborative notes, but also help combat the instability of helpfulness estimates from hyperactive minorities \cite{nudo2026hyperactive}. 

\subsection{Limitations}
\label{sec:limitations}
This work presents a study of an evolving system, that is ongoing active development. While our results reflect the current state of collaborative notes and offer opportunities for improvement, as these systems evolve over time, our findings may change. Our evaluation of the relative performance of Collaborative Notes is based on two metrics, helpfulness and time to consensus. Future work can expand the scope of this evaluation by focusing on other metrics, such as the effects of collaborative notes on post engagement, using methodology developed in \citet{slaughter2025community}. Similarly, another extension of our work would consider the relative difficulty of posts that attract collaborative notes compared to non-collaborative posts, using methodology developed in \citet{wack2026laziness}, further shedding light on the potential of collaborative notes to improve overall coverage of posts that receive notes.

\section{Acknowledgments}
We thank Alexandros Efstratiou and Haiwen Li for their detailed and valuable feedback on earlier versions of this manuscript.

\noindent
\textbf{AI disclosure:} We used Claude Code as a general-purpose coding assistant to refine figures and verify data analyses. We also used Claude Sonnet 4.6 as a grammar- and spell-checking tool and to solicit feedback on the clarity of the text. We assume full responsibility for all content produced in connection with this work.

\section{Ethics Statement}
\label{app:ethics}
All data used in this study is publicly available and fully anonymized. Contributor IDs used in our analysis cannot be traced back to individual X profiles. We adhered strictly to the policies and terms of service of the X platform and handled all collected data with respect for user privacy. While we will publicly release replication code to support reproducibility, we do not believe it can be repurposed towards harmful ends. The University of Washington IRB further determined that this work does not constitute human subject research.

\bibliography{aaai2026}

\appendix

\section{Appendix}

\subsection{Performance of Collaborative Notes across Topics}
\label{appendix:topics}

\begin{figure}[h]
    \centering
    \includegraphics[width=\columnwidth]{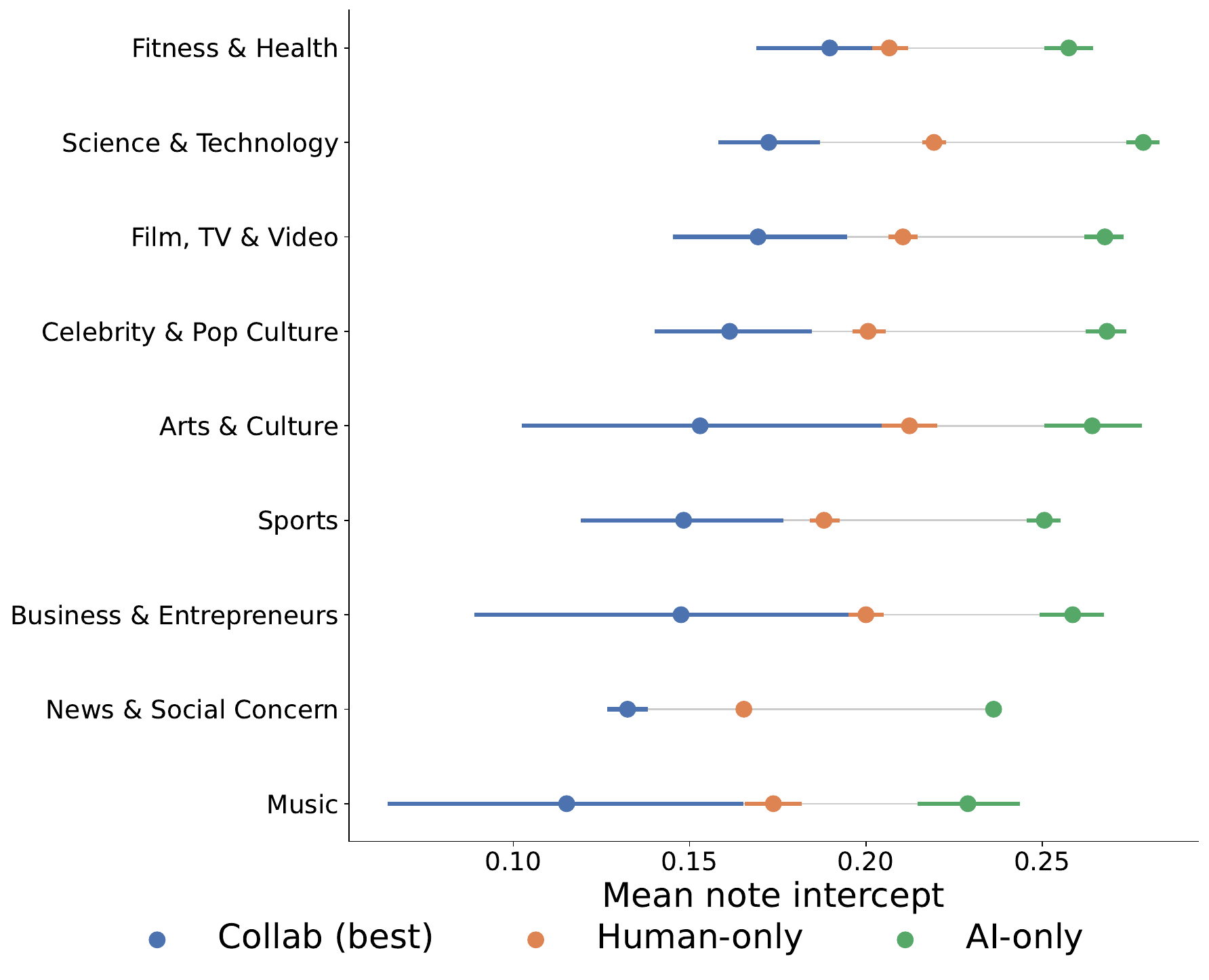}
    \caption{Collaborative notes currently under-perform non-collaborative notes consistently across all topics. Only topics with at least 20 scored collab (best) notes shown. Error bars represent bootstrapped 95\% CIs.}

    \label{fig:collab_vs_others_topics}
\end{figure}

To validate topic interpretation, we present below, for the ten topics used in Figure \ref{fig:collab_vs_others_topics}, the 15 words most over-represented per the $\chi^2$ statistic in the posts of each topic compared to the overall corpus. We note that our topic model \ref{sec:methods} yields 19 topics, of which we only retain the 10 most commonly appearing topics. 

\begin{description}
    \item[\textit{Arts \& Culture} ($n=3{,}625$)]
    \textit{art, artist, artwork, painting, jesus, ancient, century, bce, pyramids, original, poem, museum, god, artworks, archaeological}

    \item[\textit{Business \& Entrepreneurs} ($n=12{,}133$)]
    \textit{company, betting, giveaway, scam, money, policies, ads, amazon, crypto, scams, affiliate, advertising, project, partnership, investment}

    \item[\textit{Celebrity \& Pop Culture} ($n=12{,}605$)]
    \textit{taylor, swift, bieber, instagram, justin, cardi, music, megan, celebrity, gala, gaga, rihanna, nicki, zendaya, stallion}

    \item[\textit{Film, TV \& Video} ($n=15{,}527$)]
    \textit{video, movie, film, ai, avengers, youtube, reel, doomsday, generated, imdb, trailer, movies, marvel, videos, netflix}

    \item[\textit{Fitness \& Health} ($n=9{,}146$)]
    \textit{cancer, health, nih, nlm, ncbi, vaccines, vaccine, study, autism, medical, pmc, diseases, pubmed, disease, fda}

    \item[\textit{Music} ($n=3{,}416$)]
    \textit{music, spotify, album, song, songs, billboard, artist, lyrics, streams, bts, albums, guitars, kworb, tour, gibson}

    \item[\textit{News \& Social Concern} ($n=199{,}989$)]
    \textit{iran, trump, news, israel, war, epstein, iranian, government, president, politics, law, police, israeli, party, military}

    \item[\textit{Science \& Technology} ($n=24{,}194$)]
    \textit{nasa, ai, moon, earth, artemis, space, science, ii, lunar, generated, mars, image, mission, artificial, solar}

    \item[\textit{Sports} ($n=20{,}567$)]
    \textit{football, sports, league, cup, game, espn, player, match, sport, players, team, nfl, nba, arsenal, soccer}
\end{description}

\begin{table*}[ht]
\centering
\caption{Annotation Taxonomy: Axes, Categories, and Counts} 
 
\label{tab:taxonomy}
\renewcommand{\arraystretch}{1.25}
\begin{tabular}{p{2.4cm} p{3.2cm} p{3.6cm} p{5.4cm} r}
\hline
\textbf{Axis} & \textbf{Category} & \textbf{Description} & \textbf{Example} & \textbf{Count} \\
\hline

\multirow{7}{2.4cm}{\textbf{Suggestion Intent}}
  & \textsc{challenge\_claim}  & Directly challenges claim(s) made in the note & \textit{``Incorrect -- the full clip shows she talked about Israel. Claiming she was referring to the 'system' was a post hoc rationalization.''} & 4191 \\
  & \textsc{meta\_cn\_policy}         & Higher level comments on program policy violations, or whether a note is needed or not & \textit{``Note not needed''}& 2236 \\
  & \textsc{add\_context}     & Suggestions the addition of missing context or request additional information & \textit{``Can you point out the difference between net worth and cash? This post seems to confuse the two. ''} & 1637 \\
  & \textsc{source\_critique}     & Critiques source quality cited in note & \textit{``More reliable sources are needed. Wikipedia is not a reliable source, contributors are unscreened and often do not have the experience or professionalism to fact-check.''} & 768 \\
  & \textsc{value\_judgment}     & Subjectively express personal opinions, judgment. & \textit{``Why are you defending a pedophile?''}& 696 \\
  & \textsc{ai\_resistance}      & Clear expression of resistance against the use of AI & \textit{``Delete the note. Stop doing AI notes. They’re for the community not a bot''} & 502 \\
  & \textsc{formatting\_edit}     & Wording, grammar, or structure edits  & \textit{``Can you format the note a little so that sources are on a new line ''} & 428 \\
  & \textsc{low\_quality}     & Suggestion text is too short, incomprehensible or incomplete & \textit{``test hi''} & 143 \\
\hline

\multirow{3}{2.4cm}{\textbf{Evidence Level}}
  & \textsc{not\_verifiable}      & No verifiable details & \textit{``Male loneliness epidemic is empirical fact. Men are in a unique crisis.''} & 5157 \\
  & \textsc{verifiable\_no\_link} & Verifiable claims (name, place, event, etc.) but without links & \textit{``Video is from Oct 2024, not 2026. Original by Shafar Delgado shows dancing on train; Shibuya. 2024 reports indicate Japanese outrage over disruption. ''} & 4071 \\
  & \textsc{has\_links}           & Includes URLs & \textit{``This was debunked by official Fortnite PR. https://x.com/Fort\dots''} & 1373 \\

\hline

\multirow{3}{2.4cm}{\textbf{Actionability}}
  & \textsc{high}                 & Specific, directly usable edit & \textit{``Neutralize the language. This is too direct. Approach this community note as factually as possible without any narrative.''} & 4347 \\
& \textsc{medium}               & Generic but usable guidance & \textit{``None of the sources show the actual picture posted in the original post. Speculating over a table isn't enough to prove a date for the pic.''} & 3352 \\
  & \textsc{low}                  & Not useful for editing (reaction/opinion) & \textit{``He’s distracting from the Epstein files.''} & 2902 \\

\hline

\end{tabular}
\end{table*}

\clearpage
\onecolumn

\subsection{LLM Annotation Prompts}
\label{app:llm-prompts}

\begin{tcolorbox}[title={Suggestion Intent (\texttt{gpt-5.5})}, breakable]
\begin{lstlisting}[basicstyle=\ttfamily\scriptsize, breaklines=true]
You are a researcher analyzing Twitter Community Notes suggestions.

Input:
You will receive rows of note suggestions. Each row has:
- noteId (String)
- suggestion (String)

The suggestion was written by a human contributor to improve, critique, or react to the note.

Task:
For EACH suggestion, infer the PRIMARY intent of the writer and assign EXACTLY ONE label from below.
- CHALLENGE_CLAIM: Objectively challenges the fact, interpretation, reasoning, or assumptions made in the note.
- ADD_CONTEXT: Adds missing context/nuance without correcting ; ask for additional information or context; 
- SOURCE_CRITIQUE: Critiques the quality/appropriateness/credibility of sources
- META_POLICY_CN: Comments on Community Notes policy/process or X platform rather than the tweet or note content (e.g., "this belongs in replies", “belongs in comments”, "should write in certain language); Explicit states: NNN, Note not needed, Note is needed
- VALUE_JUDGMENT: Subjectively express personal opinions, moral/political judgment, insults; introduce off-topic or unrelated information.
- FORMATTING_EDIT: copyediting or writing/structure changes to the note itself (wording, grammar, tone, conciseness)
- AI_RESISTANCE: explicitly rejects or objects to AI-generated notes, labels, summaries, or moderation (e.g., “no AI notes,” “AI should not write this,”)
- LOW_QUALITY: only if the suggestion is not understandable or not in complete sentence

Output format (STRICT):
Return ONLY valid JSON in exactly this shape:
{
 "rows": [
   {
     "tweetId": "...",
     "noteId": "...",
     "SuggestionIntent": "..."
   }
 ]
}
Do not output any additional text.

\end{lstlisting}
\end{tcolorbox}

\begin{tcolorbox}[title={Evidence Level (\texttt{gpt-5.5})}, breakable]
\begin{lstlisting}[basicstyle=\ttfamily\scriptsize, breaklines=true]
You are a researcher analyzing Twitter Community Notes suggestions.

Input:
You will receive rows of note suggestions. Each row has:
- noteId (String)
- suggestion (String)

The suggestion was written by a human contributor to improve, critique, or react to the note.

Task:
For EACH suggestion, classify the level of verifiable supporting information and assign EXACTLY ONE label from below.

- HAS_LINKS: includes URLs such as http, https, or www.
- VERIFIABLE_NO_LINK: contains no URL, but includes specific checkable details such as names, organizations, places, dates, numbers, or identifiers.
- NOT_VERIFIABLE: opinion/reaction with no checkable details. 


Output format (STRICT):
Return ONLY valid JSON in exactly this shape:
{
 "rows": [
   {
     "tweetId": "...",
     "noteId": "...",
     "Evidence_Level": "..."
   }
 ]
}
Do not output any additional text.


\end{lstlisting}
\end{tcolorbox}

\begin{tcolorbox}[title={Actionability Level (\texttt{gpt-5.5})}, breakable]
\begin{lstlisting}[basicstyle=\ttfamily\scriptsize, breaklines=true]
You are a researcher analyzing Twitter Community Notes suggestions.

Input:
You will receive rows of note suggestions. Each row has:
- noteId (String)
- suggestion (String)

The suggestion was written by a human contributor to improve, critique, or react to the note.

Task:
For EACH suggestion, classify how actionable the feedback is for improving the note and assign EXACTLY ONE label from below.

- HIGH: Provides specific, directly actionable revision instructions. The suggestion clearly tells what should be added, removed, rewritten, clarified, sourced, or rephrased in the note.
- MEDIUM: Identifies a meaningful issue in the note but does not provide precise revision instructions. The feedback is usable but requires interpretation or additional work to implement.
- LOW: Provides little or no usable guidance for improving the note. Mostly emotional reaction, political opinion, insults, vague disagreement, or off-topic commentary.

Output format (STRICT):
Return ONLY valid JSON in exactly this shape:
{
 "rows": [
   {
     "tweetId": "...",
     "noteId": "...",
     "Actionability": "..."
   }
 ]
}
Do not output any additional text.

\end{lstlisting}
\end{tcolorbox}

\subsection{Suggestion Incorporation}
\label{app:grok_prompt}
Note that the following prompt has been copied verbatim from \url{https://github.com/twitter/communitynotes/blob/main/collaborative-note-generator/prompts.py}\\

\begin{tcolorbox}[title={Suggestion Incorporation (\texttt{grok-4.20-reasoning})}, breakable]
\label{app:llm-prompt-incorporation}

\begin{lstlisting}[basicstyle=\ttfamily\scriptsize, breaklines=true]
You are a helpful assistant that evaluates whether a single suggestion from a user is incorporated into a new version of a response.

Here is the previous version of the response:
```Previous version:```
{live_note_version_to_str(previous_live_note_version)}
```

Here is the new version of the response:
```New version:```
{live_note_version_to_str(new_live_note_result)}
```

Here is the suggestion to evaluate. A user suggested this on the previous version of the response.
Your task is to evaluate whether the new version of the response incorporates this suggestion,
primarily considering the PROPOSED_NOTE field, but also considering the other fields if the 
suggestion particularly pertains to one of the other fields:
```Suggestion ID: {suggestion.suggestion_id}```
{suggestion.suggestion_text}
```

Return your evaluation of whether the suggestion is incorporated into the new version in 
<INCORPORATED> tags. The possible values are "YES", "NO", or "PARTIALLY".
The new version will be likely different from the previous version in multiple ways; you should
only respond with "YES" or "PARTIALLY" if the new version is different from the previous version
in a way that was specifically suggested in the suggestion. Only respond with "YES" if new version
fully incorporates the suggestion. Default to "NO" if in doubt.
Also give an explanation of why you made your decision in <INCORPORATED_EXPLANATION> tags.

Example output:

<INCORPORATED>YES</INCORPORATED>
<INCORPORATED_EXPLANATION>
Example explanation
</INCORPORATED_EXPLANATION>
\end{lstlisting}
\end{tcolorbox}

\end{document}